# RDCI: A novel method of cluster analysis and applications thereof in sample molecular simulations

Theophanes E. Raptis[1], Vasilios E. Raptis[2]

[1] *Computational Applications Group, Division of Applied Technologies, National Centre for Science and Research "Demokritos", Athens, Greece*

[2] *Chemical Engineering Department, Imperial College London, London, England, UK*

**Abstract** A novel method, termed Reduced Dimensionality Cluster Identification, RDCI, is presented, for the identification and quantitative description of clusters formed by $N$ objects in three dimensional space. The method consists of finding a path, as short as possible, connecting the objects, and then tracking down the size $s$, of a subgroup $i-n$, $i-n+1,...$ , $i+n$, of $2n+1 < N$ particles for $i$ varying from $n+1$ to $N-n$. Clusters are located where local minima of $s(i)$ occur whereas local maxima serve as delimiters partitioning the path in subsets containing the clusters. Minimal post-processing allows for the removal of outliers on the basis of user-defined criteria and the identification of clearly defined clusters. The advantage of the method is that it requires no predetermined input or criteria of "clusterness" such as number of objects or size of aggregates. Among the numerous possible applications of the method, results are herein reported, from Molecular Dynamics simulations of a binary mixture of Lennard-Jones fluids and of a model polymeric system with a gas-like substance dispersed in it. The method is shown to allow the extraction of meaningful quantitative information regarding the tendency of molecules to cluster or de-cluster and the dynamics of the clustering processes.

**1. Introduction**

The problem of cluster analysis is well known in the statistics community and has affected numerous other branches of science wherever one has to rely on experimental or numerical methods producing vast amounts of raw data. A variety of algorithms has been created to this aim most of which contain implicit assumptions about the clusters' shape and other characteristics or require some predetermined free parameters, to work out a solution [1]. These algorithms differ also in the exact definition of what constitutes a cluster [2], a fact that is also related with the famous "Sorites Paradox" in classical (Aristotelian) logic [3].

A similar problem of locating and classifying clusters at the microscopic scale of matter arises in numerous applications involving the study of molecular systems either experimentally or via the method of numerical simulation. For instance, cluster formation may take place wherever phase separation phenomena occur while a system of interacting particles reaches an equilibrium state. The particular problem then constitutes a special case of cluster analysis in a three dimensional space assuming a direct coordinate representation of molecular positions for the produced data. In order to efficiently and effectively solve this problem a novel algorithm is introduced herein, characterised by a minimal set of free input parameters and having the additional advantage of not assuming any special definition of what constitutes a cluster.

The success of the method introduced is based on a special technique that leads to an ultimate "Dimensionality Reduction" [4], hence, the name Reduced Dimensionality Cluster Identification, RDCI, is proposed. This is achieved by introducing a carefully selected path among the particle coordinates that renders the problem effectively one-dimensional. The production of such a path has as a prerequisite the solution of a TSP ("Travelling Salesman") problem [5], which is generally considered to be NP-Hard. Nevertheless, in the method presented it is not necessary to locate a unique optimal solution but only an approximate one. This is similar to saying that an ensemble of paths is sampled and an approximate solution to the TSP problem is picked up so that the method converges rapidly without this affecting the final clustering.

Besides its simplicity and ease of implementation, another important advantage of our method over existing ones, is that it does not require any presumptions regarding the spatial size and shape of clusters, the number of particles in them, the interparticle distances or any other characteristics. RDCI utilizes the size of a subset of the examined data. However, this is merely an auxiliary quantity that can be easily optimised as will be shown in the subsequent examples, and is not directly connected with the clusters' size. The clusters correspond to topological characteristics of a geometrical object, (the aforementioned path) formed by the entire system under consideration; they are, thus, not correlated to any a priori numerical bounds of specific properties like number of clusters or size or number of particles clustered.

In this article we are showing the applicability of RDCI in three-dimensional problems by considering applications in the area of molecular simulation. In particular, in Section 2, the principles of the method are explained and the particular stages (dimensional reduction, cluster identification, refinement) that are necessary parts of its implementation, are described. In Section 3, specific applications in the analysis of clusterisation processes in binary atomistic and polymer/gas mixtures are presented and discussed. Finally, in Section 4, some conclusions are drawn and future goals and prospective applications are briefly considered.

**2. Principles and implementation of the method**

The method herein presented is inspired by the technique of Raptis *et al* [6] for the detection of diffusive jumps of small molecules in polymer matrices. In it, the trajectory of a diffusing particle is recorded at regular time intervals resulting in a series of $N$ successive positions in space. If a subset of $2n+1 < N$ positions "slides" from the beginning to the end of the recorded trajectory and its size is tracked down, e.g. in the form of a computed radius of gyration, $R_g$, it is found that the occurrence of particle jumps can be identified with the local maxima of the $R_g(t)$ curve. This is a direct consequence of the simple fact that a jump from one site to another distant one in space results in the spreading of successive particle positions over larger regions. On the contrary, when the particle spends a long time in cavities formed by surrounding molecules of the polymer matrix, the size of the set of $2n+1$ successive positions is smaller and a local minimum instead of a maximum is observed.

The present method is inspired by the latter fact and in particular the observation that local minima in the $R_g(t)$ curve correspond to more densely populated regions in space, i.e. clusters of successive particle positions. So, the present method relies on exactly the same principle as the diffusion jumps detection technique, i.e. tracking down the size of a subset of particle positions from a given data series, but the way it is applied differs in two aspects: firstly, it makes use of the principle in the inverse manner, looking for minima rather than maxima; secondly, a cluster analysis problem does not, in principle, contain the parameter of time (the data set represents positions of different particles recorded simultaneously instead of successive positions of one and the same particle). Therefore, a path has to be defined somehow.

Our implementation consists of three successive stages: the setup or dimensional reduction stage, the cluster identification stage in which the core concepts of the method are applied and the post-processing stage. This is not an arbitrary division; it emerges by the very nature of the method: one has to determine a short enough path connecting the particles by sampling the solution space of a generic TSP problem (setup), partition it according to the aforementioned criterion of local minima (cluster identification) and then apply whatever criteria better suit the problem at hand to post-process the information obtained. The essential part is the cluster identification as will be explained in the next paragraphs.

If we consider a set of $N$ objects arranged in three-dimensional space, in the first, setup, stage, a path is created that connects each particle position with the next one using a nearest-neighbors technique (in our simulations presented in the next section, periodic boundary conditions are of course taken into account). Ideally, one would search for an optimal solution to the Travelling Salesman Problem. Actually, this is not strictly necessary. For reasons that will be clear in the next paragraph, it is sufficient if the path scans the volume occupied by a particular cluster before going to the next one. Therefore, the particular method does not search for a minimal path and only filters out one that does not oscillate between neighbouring clusters. In our implementation, a simple scheme was chosen, in which all particles are successively employed as starting points and, for each one, the nearest neighbour of object *i* is chosen as the (*i*+1)-th point. Among all paths thus constructed, the shortest is picked and further refined by a simple subroutine which repeatedly exchanges randomly selected point pairs until a minimal path length is reached within a predetermined number of iterations. An example of such a path for a trial system of Lennard-Jones molecules is shown in Fig. 1.

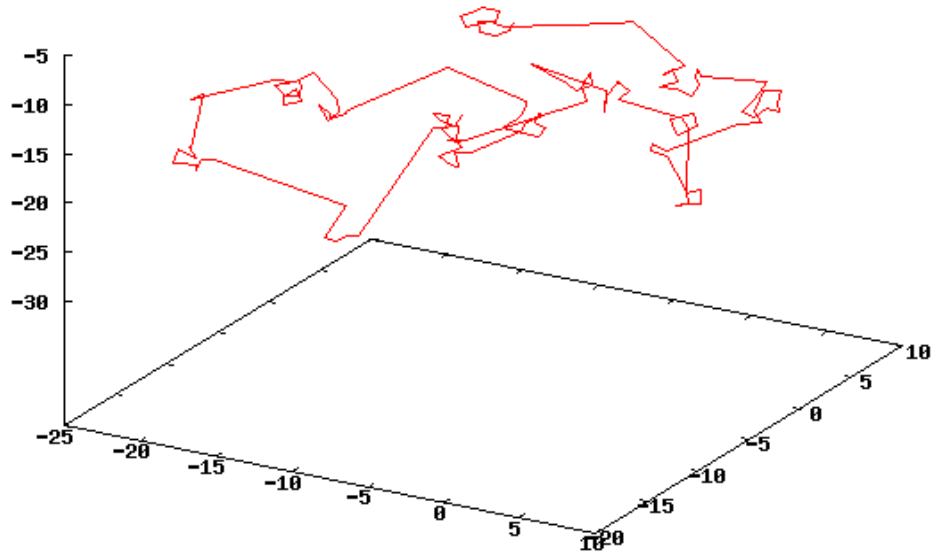

**Figure 1**. A path connecting the clustered particles. Clusters are clearly visible as "blobs" of points along the path. Size estimators for subsets of a given number of particles defined along the path, would result in smaller values (local minima) for those blobs.

At the identification stage, a "walker" representing a serial scanning of the topologically one-dimensional object just created, computes the radius of gyration, $R_g$, as well as the overall gyration tensor [7] over a set of nearest neighbour path points. In other words, a loop is performed over all points from $i = n+1$ to $N-n$, and the radius of gyration of the subset {*i-n*, *i-n*+1,... ,*i+n*} as a function of the objects' positions, $r_i$,

$$R_g(i) = \sqrt{\frac{\sum_{j=i-n}^{i+n}(\mathbf{r}_i - \mathbf{r}_{cm})^2}{2n+1}}, \quad \mathbf{r}_{cm} = \frac{\sum_{j=i-n}^{i+n}\mathbf{r}_i}{2n+1} \qquad (1)$$

is tracked down with increasing *i*. The neighbourhood radius, *n*, representing the scanning window is preselected by the user at input. The outcome is a set of values defining a $R_g$ curve over the path points. Location of successive minima immediately allows identification of each individual cluster. Neighbouring local maxima also define the edges of each cluster that is the path points at which a specific cluster "starts" or "ends". We choose to call this a collection of primitive aggregates. The identification stage constitutes the core of our method and the underlying principle is illustrated in Figure 2. It should be made clear, at this point, that the neighbourhood radius, $n$, does not constitute a measure of the clusters' size. It is merely an auxiliary parameter that helps obtain a $R_g$ curve. Sizes of identified clusters may be distributed over a certain range. The method performs better if $2n+1$ is close to the average of that distribution, but it does not have to coincide with it.

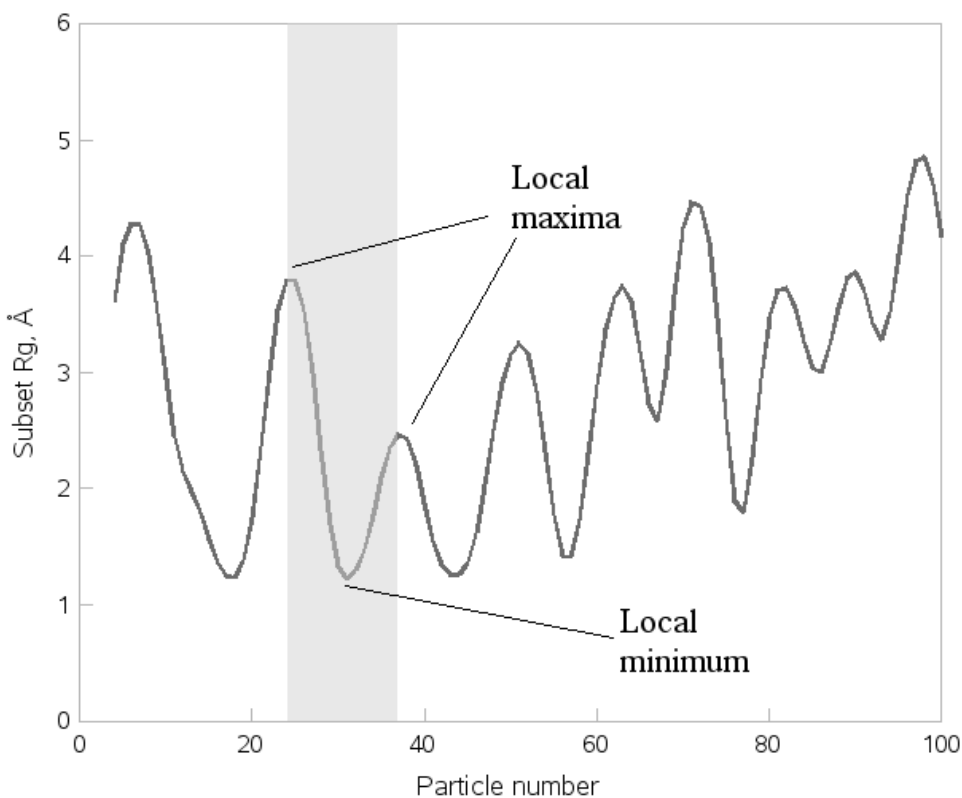

**Figure 2**. Illustration of the underlying principle of the method. The graph shows the change of the rolling subset's radius of gyration with number of the subset's central particle; the shaded area contains a primitive aggregate located at the minimum shown in the figure and delimited by the two adjacent local maxima.

At the post-processing stage, we have the freedom to apply a number of criteria that other existing cluster analysis methods cannot separate from the main body of the algorithm. In our test cases, for each set of primitive aggregates, a further refinement is introduced consisting of a direct measurement of the pair of distances between each point of a given aggregate's centre of mass and that of each neighbouring one. This allowed applying certain corrections corresponding to misclassification events such that all particle positions to be attributed to the most close aggregate. Further on, a user-defined cutoff distance is introduced such that outliers are removed, defined as particles situated farther than the cutoff with respect to the cluster's center of mass.

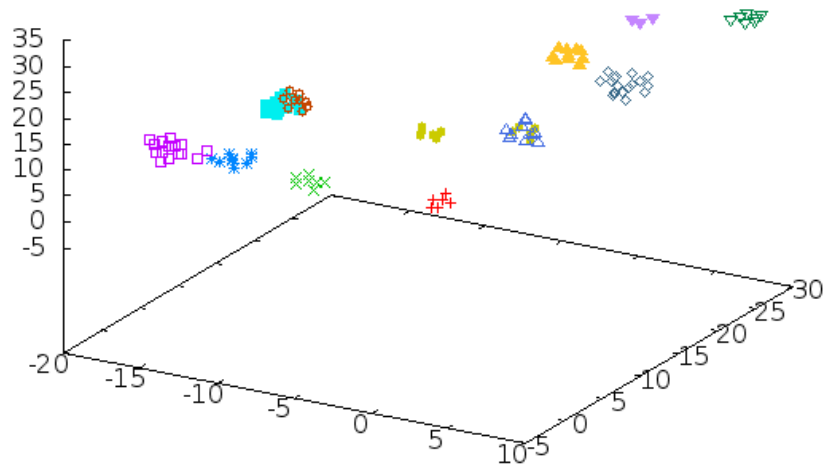

**Figure 3**. A set of clusters determined via the RDCI method. Some points or whole clusters have been discarded by means of user-defined post-processing criteria.

When one outlier is removed, the centre of mass and $R_g$ are recomputed and the procedure is repeated until convergence. At present the cutoff used has been adjusted at a value of $2R_g$. The final outcome is the refined collection of clusters. Any other set of criteria is applicable at this point in order to farther filter the final cluster distribution. An additional criterion used heretofore as explained in the next section is one of density, in order to make a proper cluster statistics according to the definitions required by the problem studied as in the exemplary case of atomistic cluster formation examined below. One such collection of clusters is shown in Figure 3.

It should be stressed that one is by no means obliged to use the specific preparation and processing techniques described in the previous paragraphs. A variety of possibilities emanates from alternative implementations of the setup and post-processing stages, so that one can talk about a family of similar techniques rather than one particular method. For instance one can choose among a variety of algorithms to connect the particles by solving the Travelling Salesman problem or incorporate the criteria of "clusterness" that best fit to the problem at hand, to refine the final results. Indeed, large sets of objects may require special techniques to efficiently generate suitable paths. This point will be further discussed in the conclusions section.

## 3. Demonstration of the Method

### *3.1 General considerations*

To demonstrate the method we chose two kinds of systems containing single atom particles dispersed among other compounds, simulated them employing the Molecular Dynamics method with the aid of the DL_POLY molecular simulation package [8], and watched the tendency of said particles to cluster. In all our simulations, the systems were initially constructed using identical parameters for the non-bonded interactions, so that no special "preference" towards specific kinds of particles was felt by each atom. Once the systems were equilibrated, the force-field parameters were altered, as will be explained in the subsequent paragraphs, so that the particles in question would feel a stronger attraction towards each other rather than their environment whereas they would tend to repel particles of the other kind(s). This way, cluster formation was forced to take place.

In all simulations, certain criteria were introduced to filter out final particle aggregates that would not merit to be considered as real clusters. In particular, only aggregates that contained at least three atoms, and at least two of them closer than $R_g$ to the cluster's center of mass were taken into account. We chose this criterion because in our systems there was a rather high fraction of the monitored species, so two atoms of the same kind colliding would not be an infrequent event worth noting. Therefore, clusters with three or more atoms, and at least two of them "close enough", were recorded.

In addition, we imposed a density constraint. In particular, we considered the "inner" region of each cluster that we shall refer to as the cluster's core, defined as a sphere of radius $R_g$ (the cluster's radius of gyration) centered at the cluster's centre of mass. Then, if $\rho = N_s / V$ is the system's number density in terms of atoms of the monitored species, then clusters that have a core number density, $\rho_c$ = {number of particles in the core}:{core's volume}, lower than $2\rho$ were discarded. Alternatively, one can use the volume of the ellipsoid defined by the gyration tensor eigenvalues [7], to compute another measure of the cluster density. This is much smaller than the spherical core volume, so its usage would constitute a weaker constraint for a given value of threshold density. Again, it should be noted that it is up to the user to decide what kind of criteria to use to filter the output.

Once the final clusters were determined, it was easy to record and average particular quantitative descriptors of their population, size and shape. These were computed for every recorded configuration and their average (where applicable) over all clusters in that configuration was recorded with time. This way, one can infer interesting conclusions about the extent and dynamics of clustering and de-clustering processes in each system. The descriptors computed in the present study included the following:

Population measures, namely

- Number of clusters per configuration

- Fraction, %, of atoms belonging to a cluster at a particular time-step. It is worth noting, by the way, that by its very nature our method constitutes a scheme of strict partitioning, i.e. each particle belongs to exactly one cluster.

Size and particle concentration estimators:

- Size of clusters in terms of radius of gyration, $R_g$.

- Number of atoms in whole clusters as well as in the clusters' "cores".

- Square roots of eigenvalues, $\lambda_x$, $\lambda_y$, $\lambda_z$, of the gyration tensor defined as [7]:

$$S_{\alpha\beta} = \frac{1}{2M^2} \sum_{i=1}^{M} \sum_{j=1}^{M} \left( r_\alpha^i - r_\alpha^j \right)\left( r_\beta^i - r_\beta^j \right) \quad (2)$$

where M is the number of atoms in the cluster and $r$ denotes the coordinates of the $i$-th and $j$-th cluster's atoms, $\alpha, \beta = x, y, z$.

- Average "core" density,

Shape descriptors, namely asphericity, acylindricity and relative shape anisotropy [7]. Assuming $\lambda_x^2 \leq \lambda_y^2 \leq \lambda_z^2$, these are defined via the following relations:

- Asphericity:

$$b = \lambda_z^2 - \frac{1}{2}\left( \lambda_x^2 + \lambda_y^2 \right) \quad (3)$$

- Acylindricity:

$$c = \lambda_y^2 - \lambda_x^2 \quad (4)$$

- Relative shape anisotropy:

$$\kappa^2 = \frac{b^2 + 3c^2/4}{R_g^4} \quad (5)$$

In our calculations we normalised the eigenvalues relative to the radius of gyration prior to computing the above quantities. Eigenvalues of individual clusters were employed to compute $b$, $c$, and $\kappa^2$ which were then, averaged over all clusters in every recorded configuration.

Of course, these are by no means the only possible cluster properties one can compute; they merely served as a way to demonstrate the wealth of information one can unveil by employing a technique as the present one that is based on no presumptions about the clusters' characteristics.

## 3.2 Simulations of Lennard-Jones fluids and polymer/gas model systems.

**Computational details**

A binary mixture consisting of 750 atoms labelled "A" and 250 atoms labelled "B" was prepared in the form of a cubic crystal structure and was allowed to melt given an initial temperature of 300 K. The particles interacted via a 12-6 Lennard-Jones potential

$$4\varepsilon_{\alpha\beta}\left[\left(\sigma_{\alpha\beta}/r\right)^{12}-\left(\sigma_{\alpha\beta}/r\right)^{6}\right] \tag{6}$$

where $\varepsilon_{\alpha\beta}$ is potential well depth, $\sigma_{\alpha\beta}$ is the atom pair effective diameter, $r$ is the atom pair distance and $\alpha, \beta$ = A or B. At the melting stage, all particles were characterised by the same force-field parameters, Table 1a. The final structure was fed as input to the next stage. In it, we aimed at forcing the B-type atoms to come closer to each other and form clearly emerging clusters. To this purpose, we changed the interaction parameters so that the A-B potential becomes more repulsive and the B-B one more attractive whereas the A-A one was left unchanged. If we rewrite the non-bonded potential in the form $A/r^{12} - C/r^6$ we can see that increasing $A$ or decreasing $C$ would render it more repulsive and vice versa. This can be achieved by increasing the atom diameters, $\sigma$. Furthermore, by increasing $\varepsilon$, we render the potential "stickier", which would be desirable if we want to force cluster formation.

We applied the above in the following manner: Diameters, $\sigma$, for AB-type atom pairs were multiplied by $2^{1/6}$ so that the zero energy distance for the new value would equal to the one of the minimum energy for the previous diameter. Likewise, BB pair diameters were reduced by the same factor. As regards potential well depths, $\varepsilon$, we require that the one for AB pairs equal the AA potential at distance $r = 2^{1/6}\sigma_{AB} = 2^{1/3}\sigma_{AA}$ which is found to be 3 $\varepsilon_{AA}$ / 4. Conversely, the BB potential well depth is attributed a value of 4 $\varepsilon_{AA}$ / 3. The same procedure was repeated four times for different temperatures, namely 250, 300, 350 and 400 K.

Finally, one more stage followed, aimed at studying the inverse, i.e. de-clustering effect, by bringing all parameters back to their initial values. This was only done for the 300 K temperature. All parameters employed at each stage are summarised in Table 1a. All simulations were conducted at the canonical ensemble, lasted 500 ps each and the time-step was 1 fs. Cubic periodic boundary conditions were employed and the simulation's cell length was equal to 37.5 Å throughout the runs, regardless of temperature. It should be mentioned that these simulations aim merely at demonstrating the application of the presented cluster analysis method; they do not aspire providing accurate predictions but only give the reader certain guidelines and ideas about what can be done in the framework of interesting problems involving clusterisation processes.

**TABLE 1**. Force Field parameters employed in the test MD simulations: (a) binary mixture of atomistic Lennard-Jones fluids (b) polymer/gas mixture.

| (a) Binary System, 75% A, 25% B | | | | | | |
|---|---|---|---|---|---|---|
| | AA | | BB | | AB | |
| | $\sigma$, Å | $\varepsilon$, kcal mol$^{-1}$ | $\sigma$, Å | $\varepsilon$, kcal mol$^{-1}$ | $\sigma$, Å | $\varepsilon$, kcal mol$^{-1}$ |
| *Melting* | 3.750 | 0.200 | 3.750 | 0.200 | 3.750 | 0.200 |
| *Clustering* | 4.209 | 0.150 | 3.341 | 0.267 | 3.750 | 0.200 |
| *De-clustering* | 3.750 | 0.200 | 3.750 | 0.200 | 3.750 | 0.200 |
| Cutoff distance | 12.5 Å throughout all simulations. | | | | | |
| (b) Polymer/gas system, 30 $C_{100}$ chains / 150 $CH_4$ molecules | | | | | | |
| | $CH_2$-$CH_4$ | | $CH_3$-$CH_4$ | | $CH_4$-$CH_4$ | |
| | $\sigma$, Å | $\varepsilon$, kcal mol$^{-1}$ | $\sigma$, Å | $\varepsilon$, kcal mol$^{-1}$ | $\sigma$, Å | $\varepsilon$, kcal mol$^{-1}$ |
| *Melting* | 3.835 | 0.164 | 3.765 | 0.239 | 3.720 | 0.091 |
| *Clustering* | 4.305 | 0.055 | 4.226 | 0.080 | 3.314 | 0.273 |

By recording the population, size and shape descriptors mentioned in Section 3.1, we were able to illustrate various aspects of the clusterisation dynamics and its dependence on external factors like temperature, as will be explained in the next subsection. Another interesting topic regarding the factors affect the extent and dynamics of clusterisation, has to do with characteristics of the each particular system like mobility of the other molecules surrounding the clustered particles. To study this effect we simulated systems of a gas-like substance dispersed in a polymer-like matrix, so that we can compare with results from the binary atomistic mixtures.

Small penetrant molecules dispersed in slowly moving polymeric materials are trapped for long time intervals in free volume cavities which they escape from time to time, thanks to thermal fluctuations. The polymer matrix slows down the penetrants diffusive motion which often takes place in the form of infrequent jumps (depending on the rigidity of the matrix). This would have an effect on clusterisation processes as well. On the other hand, the presence of free volume cavities where penetrant molecules can be hosted implies that there might be a natural tendency of the low molecular weight species to aggregate in those regions (depending also on the energetics of their interaction with the polymer), which would compensate for the slowing down due to the polymer relative rigidity.

To study the above effects we constructed a system of 30 polyethylene-like chains each consisting of 100 united atoms, namely 98 methylene groups and 2 methyl groups capping the chain ends. In it, 150 methane-like united atom particles were randomly dispersed. Cubic periodic boundary conditions were imposed and the simulation box length was equal to 46.422367 A. The initial structure was generated in the same way as described in [6] and [9], which is briefly outlined here: All penetrant molecules and the first atom of each polymer were randomly placed in the simulation box. Placing the methane molecules inside the box from the very beginning ensured that they would be rather homogeneously distributed in space, not being obstructed by the polymer chains. Then, polymer molecules were built in a step-wise manner by adding one bond to each molecule at a time. Bond orientations were chosen, among a set of trial directions, using a Metropolis criterion and rejecting configurations with non-bonded atom pairs closer than a hard-sphere diameter of 2.5 A. Then, the system's potential energy was reduced by slightly displacing the atoms in various random directions. This was repeated with atom diameters growing from the hard-sphere diameter to their normal values and the relaxation process was over when smooth and realistic distributions of angles and torsions were obtained.

The force-field parameters were the ones of the well-knwon NERD model [10] for linear alkanes. First, a 6 ns MD simulation was conducted, allowing for a sufficient equilibration of the system. The cluster shape and size descriptors were calculated and were seen to fluctuate around their mean values throughout the simulation (not reproduced here). Then, the non-bonded potential sigma parameters were altered in the same way as in the case of the binary atomistic mixture, to force clustering of methane-like molecules. Because of their broadly distributed values in the original NERD representation, epsilons were altered by an arbitrarily chosen factor of 3 or 1/3, depending on the pair type, to ensure stronger attractive or repulsive interactions for the gas-gas and gas-polymer atom pairs, respectively. These parameters are presented in Table 1b. For the rest of the nonbonded parameters as well as the bonded ones, the reader is referred to the NERD force-field original publication [10].

**Results and discussion**

In all cluster analysis computations, a neighborhood radius, $n = 4$, was used. In certain cases, $n = 5$ or 6 was used to ensure invariability of results. The setup, cluster identification and post-processing stages were repeated over all recorded configurations of every MD trajectory and the size and shape descriptors listed in Section 3.1 were computed and recorded.

Transforming the potential in the way described in the previous subsection, allowed for the formation of many new clusters within a time period of about 200 to 300 ps, as evidenced by a substantial increase in the aforementioned population measures. Then, the system reached a state of equilibrium with all descriptors fluctuating around their mean values. In particular, the number of clusters showed a marked increase from initial values close to 5 to ones fluctuating around 20, Figure 4. Temperature did not affect this process substantially in the sense that a steady state seemed to set in at around 200 to 250 ps, although one can notice a slightly faster rate at higher temperatures.

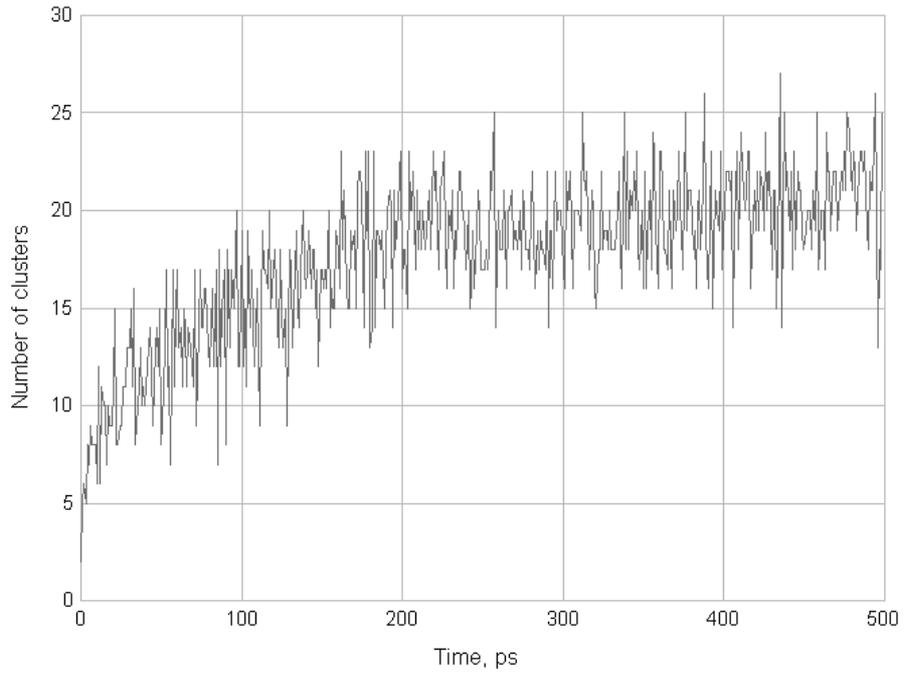

(a)

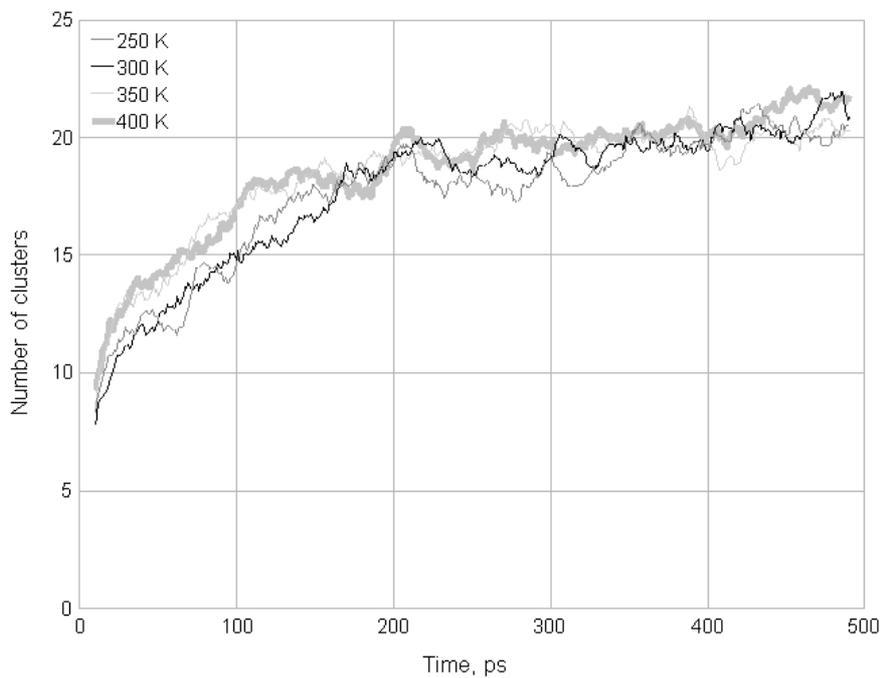

(b)

**Figure 4** Number of clusters of B atoms with time, in a 75% A / 25% B mixture of Lennard-Jones fluids. Force-field parameters have been deliberately chosen so as to force the components to separate. (a) Results at T = 300 K (b) Comparison of results at various temperatures (rolling averages over 20 data points are shown instead of the original data series, for the sake of clarity).

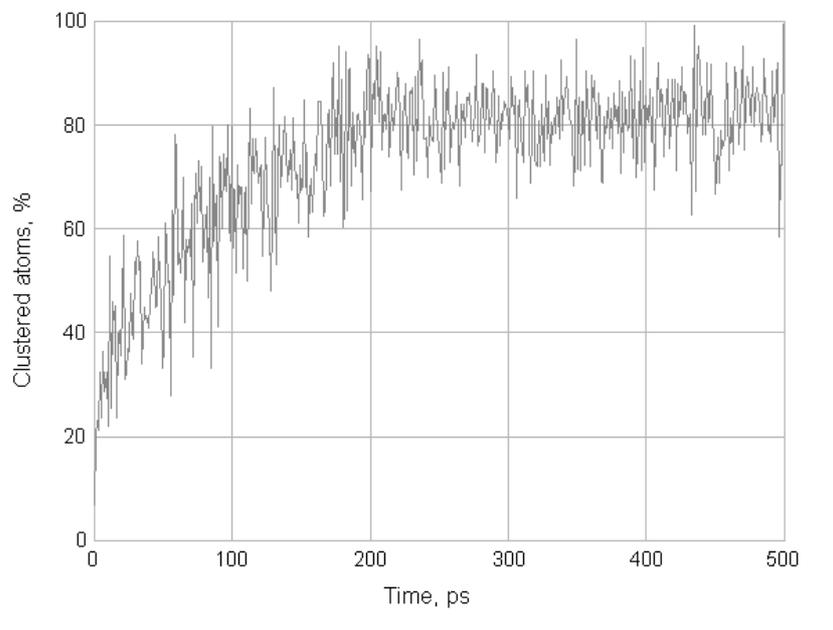

(a)

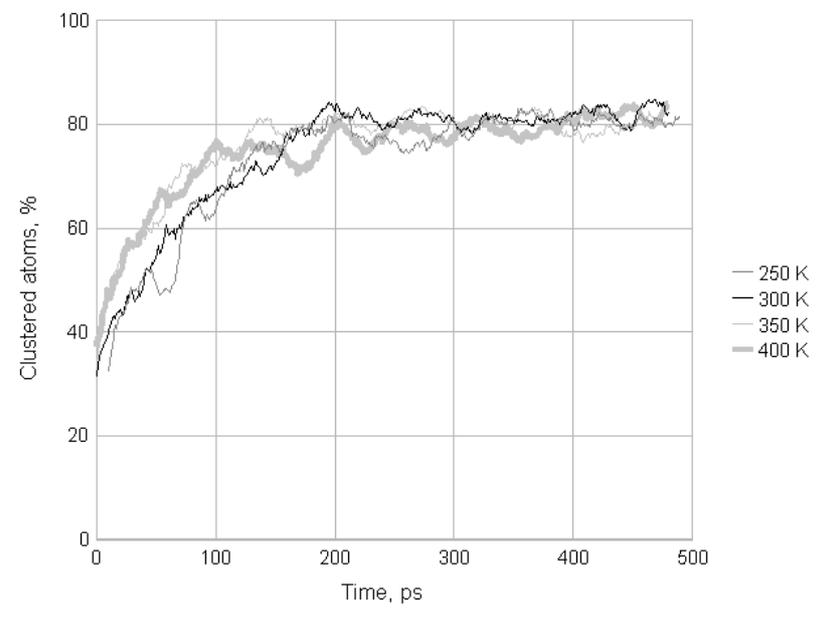

(b)

*Figure* **5** Fraction, %, of clustered B atoms with time, in a 75% A / 25% B mixture of Lennard-Jones fluids. (a) Results at T = 300 K (b) Comparison of results at various temperatures (rolling averages over 20 data points are shown instead of the original data series, for the sake of clarity).

Interestingly, we can observe two regimes in the growth stage: one that can be fitted relatively well to an exponential expression of the form $N_0 (1 - e^{-t/\tau})$, lasting until ~50 ps, and a more or less linear one (plus a strong noise), $N_0 + t/\tau$, from 50 to about 200 ps. The exponential regime parameter $N_0$ increases with temperature from approximately 12 to 14 whereas the corresponding timescale, $\tau$, is close to 8 ps. The linear regime time scale falls in the range of 16 to 20 ps (decreasing with temperature). Since we are restricted in demonstrating the capabilities of our method rather than conducting a systematic research on the clusterisation processes, we merely note this finding as a possible area to be further studied.

Likewise, the % fraction of clustered atoms, Fig. 5, increased from approx. 20% to a plateau of 80% within the same time interval as the number of clusters. Again, a slightly higher rate of clusterisation is observed at higher temperatures. A linear regime is not clearly observed regarding this descriptor. The clusters' average density also exhibits a clear increase followed by stabilisation at values between 4 and 4.5 relative to the system's number density in B-type atoms. This trend can be seen as the result of a decrease in the average number of gyration from 4.5 to 4 Å counterbalancing a slight but observable decrease in number of atoms per cluster, generally fluctuating close to 10. The establishment of a steady-state regime is also evidenced by shifted peaks and narrower widths of descriptor distributions corresponding to specific time intervals, as the fraction histograms in Figure 6 show.

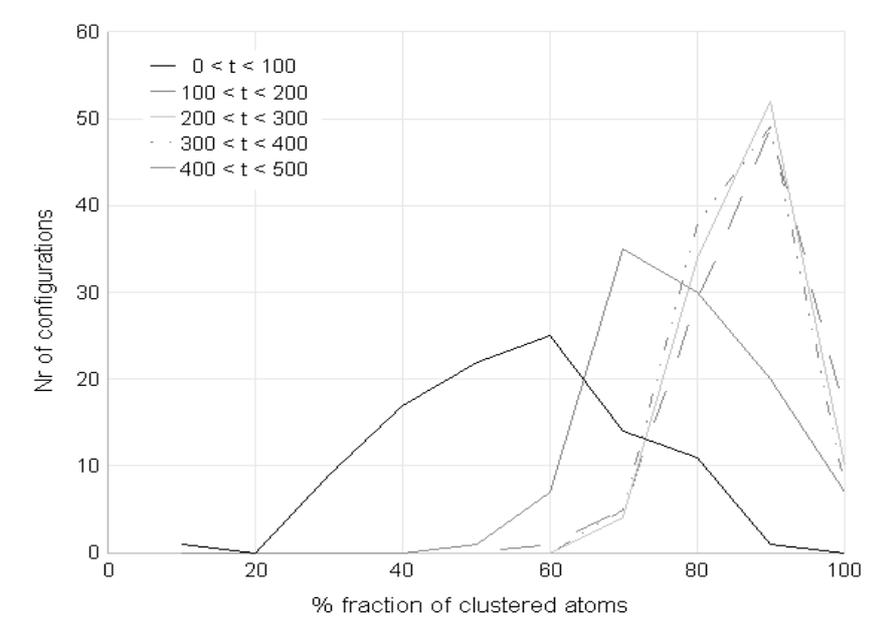

***Figure* 6** Histogram of % fraction of clustered atoms of type B in a 75% A / 25% B binary mixture of Lennard-Jones fluids at a temperature of 300 K. The distributions correspond to different time intervals. Force-field parameters have been deliberately chosen so as to force the two components to separate resulting in a clear shift of the peaks towards higher fractions as time advances. A steady state seemingly sets in at time scales beyond 200 ps.

Bringing the force-field parameters back to their initial values is seen to result in the inverse process of de-clustering, as exhibited in clearly falling tendencies of both the number of clusters and the % fraction of clustered atoms, Fig. 7. The duration of the transient regime and the kind of dynamics are quite similar for both the fraction of clustered atoms and the number of clusters. In all cases herein studied, whether clustering or declustering, a steady state seems to set in at about 250 ps. At the same time, there are certain quantities, and mainly the ones related to the clusters' shape (gyration tensor eigenvalues, asphericity, acylindricity, shape anisotropy) that remain more or less stable throughout the transient and steady state stages of our simulations.

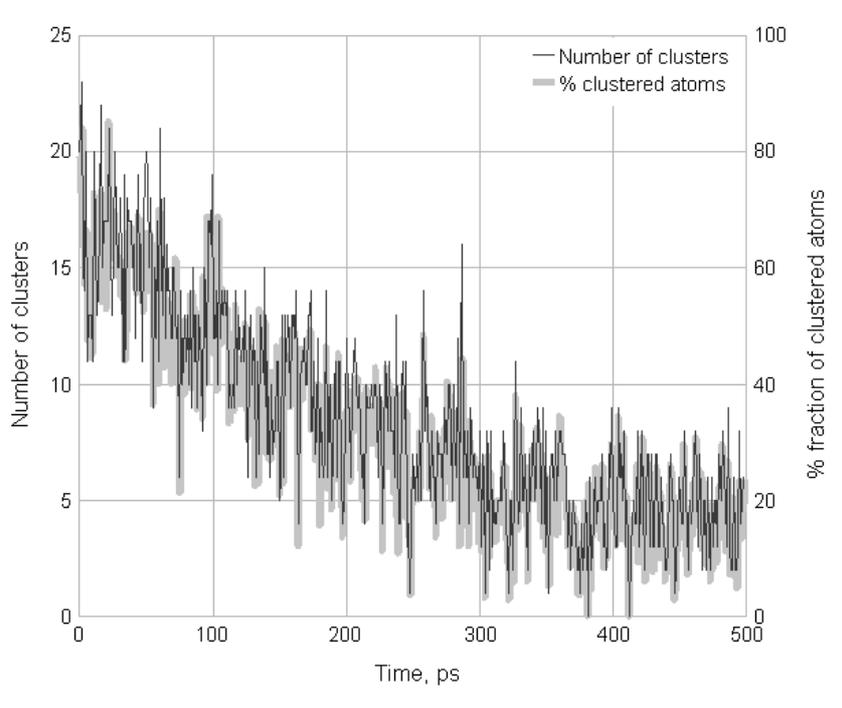

**Figure 7** Number of clusters (left vertical axis) and fraction, %, of clustered atoms (right vertical axis) of type B with time, in a 75% A / 25% B binary mixture of Lennard-Jones fluids at a temperature of 300 K. Force-field parameters have been deliberately chosen so as to force the B-type atoms to de-cluster and the two components, A and B, to mix.

In Table 2 values of all population, size and shape descriptors averaged over the last 250 ps are reported, for all simulations of the model binary system. A maximum of about 80% of the atoms are eventually clustered at all temperatures studied. The total number of clusters increases weakly with temperature, whereas the number of atoms per clusters decreases slightly. As the radius of gyration also exhibits a tendency, albeit weak, to decrease with temperature, the outcome is higher density at lower temperatures. Apparently, thermal motion of atoms prevents the clusters from growing beyond a certain point, allowing however, the formation of more clusters in dynamic equilibrium with escaping "free" atoms. Finally, all shape descriptors exhibit a consistent tendency towards irregular shapes generally classifiable as triaxial ellipsoids.

**TABLE 2**. Size/shape descriptors of B-type clusters in 75% A/25% B binary systems and polymer/gas systems, averaged over the configurations that sample the steady-state trajectories, at various temperatures. Numbers in italics denote standard deviations.

$N_{cl}$ = number of clusters, % $F_{cl}$ = % fraction of clustered atoms, $N_{at}$ = number of atoms per cluster, $N_{at,\,core}$ = number of atoms in cluster's "core", $R_g$ = radius of gyration (Å), $\rho_{rel}$ = number density of cluster's "core" relative to the system's density with respect to the considered atom type, $\lambda_{\alpha\alpha}$ = square root of gyration tensor eigenvalue ($\alpha = x, y, z$), $b$ = asphericity, $c$ = acylindricity, $\kappa^2$ = relative shape anisotropy.

| $T$, K | Population | | Size measures | | | | Shape descriptors | | | | | |
|---|---|---|---|---|---|---|---|---|---|---|---|---|
| | $N_{cl}$ | % $F_{cl}$ | $N_{at}$ | $N_{at,\,core}$ | $R_g$ | $\rho_{rel}$ | $\lambda_{xx}$ | $\lambda_{yy}$ | $\lambda_{zz}$ | $b$ | $c$ | $\kappa^2$ |
| *Binary atomistic, clustering* | | | | | | | | | | | | |
| 250 | 19.47 | 79.78 | 10.34 | 5.90 | 4.13 | 4.36 | 1.31 | 2.07 | 3.28 | 0.54 | 0.42 | 0.48 |
| | *2.60* | *7.59* | *1.05* | *0.70* | *0.22* | *0.39* | *0.10* | *0.11* | *0.20* | *0.03* | *0.05* | *0.05* |
| 300 | 19.83 | 81.34 | 10.34 | 5.84 | 4.14 | 4.25 | 1.32 | 2.09 | 3.28 | 0.53 | 0.43 | 0.48 |
| | *2.51* | *6.87* | *0.95* | *0.63* | *0.20* | *0.35* | *0.09* | *0.11* | *0.18* | *0.03* | *0.05* | *0.05* |
| 350 | 20.05 | 80.32 | 10.09 | 5.69 | 4.11 | 4.25 | 1.30 | 2.07 | 3.26 | 0.54 | 0.43 | 0.48 |
| | *2.38* | *6.83* | *0.90* | *0.57* | *0.20* | *0.36* | *0.09* | *0.12* | *0.18* | *0.03* | *0.05* | *0.05* |
| 400 | 20.47 | 80.05 | 9.85 | 5.56 | 4.09 | 4.23 | 1.27 | 2.07 | 3.25 | 0.54 | 0.44 | 0.50 |
| | *2.43* | *6.74* | *0.91* | *0.59* | *0.20* | *0.35* | *0.09* | *0.11* | *0.18* | *0.03* | *0.04* | *0.05* |
| *Binary atomistic, declustering* | | | | | | | | | | | | |
| 300 | 5.39 | 19.03 | 8.80 | 5.25 | 4.47 | 2.97 | 1.37 | 2.26 | 3.55 | 0.55 | 0.45 | 0.51 |
| | *2.39* | *8.52* | *1.67* | *1.05* | *0.43* | *0.43* | *0.20* | *0.26* | *0.36* | *0.07* | *0.09* | *0.10* |
| *Polymer/gas, clustering* | | | | | | | | | | | | |
| 300 | 12.69 | 87.28 | 10.43 | 6.02 | 5.18 | 7.65 | 1.53 | 2.54 | 4.19 | 0.57 | 0.46 | 0.55 |
| | *1.59* | *7.04* | *1.17* | *0.77* | *0.34* | *0.94* | *0.16* | *0.20* | *0.30* | *0.04* | *0.06* | *0.07* |

In Table 2, average values of the corresponding descriptors for the polymer / gas system are also included. These are block-averaged over the last 5 ns of the trajectory of the, second, forced clustering, simulation, divided in 1 ns-blocks. A stronger tendency towards the formation of clusters is observed, as evidenced by the higher fraction of clustered atoms (87% as compared to 80% for the binary atomistic systems). This is probably due to the contribution of the polymer matrix hosting effect, namely the tendency of the gas molecules to be trapped in free volume cavities for as long as the latter exist. Interestingly, it was observed that the clusterisation process takes place at an equally fast rate as in the case of the binary atomistic mixture; a time period of 200 – 300 ps is sufficient for a steady state to set in. A slightly higher number of atoms per clusters are observed but the average radius of gyration is also larger. Shape descriptors indicate that the clusters formed in the presence of the polymer are somewhat more elongated than those in the atomistic mixture, a fact not inconsistent with the free volume hosting effect. Indeed, free volume cavities are likely to be elongated or distributed along the chains, having an analogous effect on the gaseous clusters.

**4. Conclusions and future work**

Clusters are often fuzzy collections of objects in the sense that they lack well defined boundaries separating them from their environment and from each other. How can we use the language of precision (mathematics and computational modelling) to locate, classify and describe quantitatively what is imprecise by its very nature? The answer lies in the fact that certain topological characteristics like local extrema exist, that are well defined and can serve as delimiters of cluster-containing regions, once the system in question has been mapped to a suitably defined geometrical object containing all necessary information.

In this spirit, our dimensional reduction approach herein presented, alleviates the aforementioned problems that are inherent in cluster analysis. From that, stems one of the method's two main advantages: one needs not feed the algorithm with a priori assumptions about the shape and size of the clusters. This, not only allows freedom to apply the most suitable post-processing criteria to filter the output, but, above all, it provides results unrestricted by any presumptions, which, therefore reflect accurately the actual state of the system, namely distribution of size and shape of clusters and other population characteristics. Then, it is permissible to process the output statistically and, furthermore, correlate its temporal changes to extract valid and insightful information about the cluster formation or annihilation dynamics. It is exactly this possibility that we demonstrated in our test-case simulations which were shown to provide reasonable and interesting results in terms of cluster population, size, density and shape, as well as in terms of dynamics.

On the other hand, certain special problems may arise in the cases of very broad cluster size distributions or very large clusters; the latter may appear as an array of successive local $R_g$-minima being interpreted as collections of smaller aggregates rather than single sets of objects. Also, large data sets require more sophisticated path construction techniques than the brute-force employed in our demonstration examples or the setup stage would be significantly slowed down. In the present method, such deficiencies can be cured due to the modular character of the algorithm allowing replacing certain parts thereof with specialised subroutines most fit for the problem at hand. For instance, solutions akin to the linked-cell neighbour list techniques employed in molecular simulations may be helpful in addressing the problem of big data sets.

At the present stage of development, similar algorithms are studied in connection with research on very large galactic clusters from astronomical data alongside with special hierarchical filtering techniques for isolating the largest clusters among a population. Other problems of interest that are currently studied include more realistic studies of gas – polymer mixtures in various temperatures below and above the glass transition temperature and the correlation between clusterisation dynamics with diffusion jumps, alcohol-water mixtures and correlation of water-water and water-hydroxyl clusters with hydrogen bonds, as well as the study of nanocomposite polymers with nanoparticles.

While our method finds application in molecular cluster identification and tracking, it is quite evident from the generic character of the underlying principle that it is extendible in a natural way into higher dimensions and certainly applicable into other branches as well, including bioinformatics, data mining and machine learning. For this to be possible, it is required that the method will prove fail-safe against some known geometrical and topological problems appearing in higher dimensional spaces, generally falling under the label of "Dimensionality Curse" [11], where spurious correlations appear that make final classification difficult and ambiguous especially in hierarchical classification problems.

**References**

1. B. S. Everitt, S. Landau, M. Leese, D. Stahl, *Cluster Analysis*, 5$^{th}$ ed., John Wiley & Sons Ltd, Chichester, UK, 2011.

2. V. Estivill-Castro, SIGKDD Explorations, **4**, 65 (2002).

3. E. Trillas, L. A. Urtubey, Applied Soft Computing, **11**, 1506 (2011)

4. P. R. Krishnaiah, L. N. Kanal, *Handbook of Statistics 2: Classification, Pattern Recognition and Reduction of Dimensionality*, Elsevier Science Pub. Co., New York, 1983

5. G. Gutin and A. P. Punnen (editors), *The Traveling Salesman Problem and Its Variations*, Kluwer Academic Publishers, New York, Boston, Dordrecht, London, Moscow, 2004

6. T. E. Raptis, V. E. Raptis, J. Samios, J. Phys. Chem. B, **111**, 13683 (2007).

7. D. N. Theodorou, U. W. Suter, Macromol. **18**, 1206 (1985)

8. I.T. Todorov, W. Smith, K. Trachenko, M.T. Dove, J. Mater. Chem., **16**, 1911 (2006)

9. T. E. Raptis, V. E. Raptis, J. Samios, Mol. Phys., **110**, 1171 (2012).

10. S. K. Nath, F. A. Escobedo, J. J. de Pablo, J. Chem. Phys. **108**, 9905 (1998).

11. Bellman, R. *Introduction to the Mathematical Theory of Control Processes*, vol. 2, Academic Press, New York, 1971.